\documentclass{PoS}

\usepackage{epsfig}
\usepackage{amsmath}
\usepackage{amssymb}

\newcommand{\I}{\mathrm{i}}  
\newcommand{\E}{\mathrm{e}}  

\newcommand{\FC}{\;,}
\newcommand{\FD}{\;.}

\title{Decay of  $\rho$ and $a_1$ mesons on the lattice using distillation}

\ShortTitle{Decay of  $\rho$ and $a_1$ mesons on the lattice using distillation}

\author{\speaker{Sasa Prelovsek}\\
         Jozef Stefan Institute and 
        Department of Physics at University of Ljubljana, Ljubljana,  Slovenia\\
        E-mail: \email{sasa.prelovsek@ijs.si}
}

\author{C.~B.~Lang\\
        Institut f\"ur Physik, FB Theoretische Physik, Universit\"at
Graz, A--8010 Graz, Austria\\
        E-mail: \email{christian.lang@uni-graz.at}
}

\author{Daniel Mohler\\
        TRIUMF, 4004 Wesbrook Mall Vancouver, BC V6T 2A3, Canada \\
        E-mail: \email{mohler@triumf.ca}
}

\author{Matija Vidmar \\
        Jozef Stefan Institute, Ljubljana, Slovenia\\
}

\abstract{We extract the P-wave $\pi\pi$  phase shift for five values of pion
relative momenta, which gives  information on the $\rho$ resonance.  The
Breit-Wigner formula describes the $\pi\pi$ phase shift dependence nicely and we
extract   $m_\rho=792(7)(8)$ MeV and the coupling $g_{\rho\pi\pi}=5.13(20)$ at
our $m_\pi=266$ MeV. We   extract the P-wave scattering length $a_{l=1}^{\pi\pi}=0.082(10)(3)~$fm$^3$
from the state with the lowest pion relative momenta.

We also determine the S-wave $\rho\pi$ phase shift for two values of relative
momenta, which provides parameters of the lowest axial resonance $a_1(1260)$.
Using the Breit-Wigner fit we extract  $m_{a1}=1.44(4)$ GeV and the coupling
$g_{a_1\rho\pi}=1.1(3)~$ GeV. From the lowest state we also extract the
$\rho\pi$ scattering length $a_{l=0}^{\rho\pi}=0.23(12)$ fm for our $m_\pi$. 

The simulation is performed using one $N_f=2$ ensemble of gauge configurations
with clover-improved
Wilson quarks. The phase shifts are determined from the lowest two
energy-levels,  which are obtained by the variational analysis with a number of
quark-antiquark and meson-meson interpolators. The correlation functions are
calculated using the distillation method with the Laplacian Heaviside (LapH)
smearing of quarks.  
}

\FullConference{XXIX International Symposium on Lattice Field Theory \\
		July 10 - 16 2011\\
		Squaw Valley, Lake Tahoe, California}

\begin{document}

\section{Introduction}

Extracting the width of a hadronic resonance $R$ from lattice QCD is challenging. 
The only proper method used up to now applies to resonances $R$ that appear 
in the elastic scattering of two hadrons $H_1H_2\to R\to H_1H_2$. First the elastic phase shift $\delta(s)$ for $H_1H_2$ scattering has to be determined from the lattice for several values of  $s=E_{CM}^2=E^2-\mathbf{P}^2$, where $E$ and  $\mathbf{P}$ are the energy and the total momentum of the $H_1H_2$ system. L\"uscher has shown that 
the energy $E$ of two hadrons in a box of size $L\simeq $ few fm provides the value of the infinite-volume elastic phase shift $\delta(s)$ at $s=E^2-\mathbf{P}^2$ \cite{luscher}. His relation between $\delta$ and $E$   for  $\mathbf{P}=0$ was generalized to $\mathbf{P}\not = 0$ in \cite{rummukainen,sharpe, rho_etmc_pos}. 
In practice, one or two lowest energy levels $E$ are extracted and a few choices of $\mathbf{P}$ are used in order to extract $\delta(s)$ at  different values of $s=E^2-\mathbf{P}^2$. 

The resulting $\delta(s)$ can be fit with a Breit-Wigner (or any other desired) form, where both are related via 
the scattering amplitude $a_l$ for the $l$-th partial wave 
\begin{equation}
\label{bw}
a_l=\frac{-\sqrt{s}\,\Gamma_R(s)}{s-m_R^2+i \sqrt{s}\,\Gamma_R(s)}=\frac{e^{2i \delta(s)}-1}{2i} \quad \mathrm{or} \quad \sqrt{s}\,\Gamma_R(s)\,\cot \delta(s)=m_R^2-s\FC \quad \Gamma_R(s)\propto g_{RH_1H_2}^2 \frac{{p^*}^{2l+1}}{s}
\end{equation}
and $p^*$ is the momentum of $H_1$ and $H_2$ in their center-of-momentum (CMF) frame.  
This relation can be used to extract the 
mass $m_R$ and the width $\Gamma_R=\Gamma_R(m_R^2)$ of the resonance 
from lattice data on $\delta(s)$. The width depends significantly on  the phase space and therefore on $m_\pi$, so it is common to extract the coupling $g_{RH_1H_2}$, which is expected to depend only mildly on $m_\pi$.  

Among all the meson resonances only the $\rho$ meson width  has been determined 
properly using this method. The first lattice determination was done by PACS-CS in 2007 \cite{rho_cppacs}. Since then, several studies of the $\rho$ have been carried out \cite{rho_qcdsf,rho_bmw}, with the most recent ones  \cite{rho_etmc,rho_our,rho_pacscs}.  In this talk we present our recent study of the $\rho$ \cite{rho_our}, 
which achieves the smallest statistical  errors (on one ensemble only, however) on the resulting $\delta(s)$, $m_\rho$ and $\Gamma_\rho$ due to several improvements listed below. 

We also extract the S-wave $\rho\pi$ elastic phase shift, which enables us to extract the mass $m_{a1}$ and the width $\Gamma_{a1}$ of the lowest lying axial resonance $a_1(1260)$. The lattice study of this resonance is especially  welcome as the experimental knowledge on it is very poor: the width has a wide range  $\Gamma_{a1}^{exp}=250-600$ MeV \cite{pdg}, and none of its branching ratios have been reliably determined \footnote{All final states are quoted just as "seen" in \cite{pdg}.}  \cite{pdg}. To our knowledge, this is the first lattice study aimed at the $\rho\pi$ scattering and $\Gamma_{a1}$. 

\section{Lattice simulation}

We use 280 $N_f=2$ configurations with  tree-level clover-improved Wilson dynamical and valence quarks, corresponding to $m_\pi a=0.1673(16)$ or $m_\pi=266(3)(3)$ MeV \cite{hasenfratz}. 
The lattice spacing  $a=0.1239(13)$ fm was determined using the Sommer parameter $r_0$ \cite{rho_our} and our  $N_L^3\times N_T=16^3\times 32$ is rather small, allowing us to use the powerful but costly full distillation method \cite{peardon}. We combine  periodic and anti-periodic propagators in time to reduce the finite $N_T$ effects \cite{rho_our}. 

\section{$\rho$ resonance and $\pi\pi$ phase shift}

The details of our lattice simulation aimed at $\pi\pi$ phase shifts and the $\rho$ resonance have  been published in \cite{rho_our}. In this talk, we emphasize the most important steps and results.

The $\pi^+\pi^-\to \rho^0\to\pi^+\pi^-$ scattering is elastic below the $4\pi$ threshold $\sqrt{s}<4m_\pi$ and we can apply L\"uscher's method. 
We determine the lowest two energy-levels of the  $\rho^0\leftrightarrow \pi^+\pi^-$ coupled system with $J^{PC}=1^{--}$ and $|I,I_3\rangle =|1,0\rangle$ for the following cases of total momentum $P$ 

\begin{table*}[h!]
\begin{center}
\begin{tabular}{c | c c c}
 $\mathbf{P}$ & group & irrep& decay\\
\hline
$\mathbf{0}$     &$O_h $ & $T_1^-$  &$ \mathbf{\rho_3}(\mathbf{0})\to \pi(\mathbf{e_3})\pi(-\mathbf{e_3})$\\
$\tfrac{2\pi}{L}\mathbf{e_3}$& $D_{4h}$&$ A_2^-$ & $\mathbf{\rho_3}(\mathbf{e_3})\to \pi(\mathbf{e_3})\pi(\mathbf{0})$\\
$\tfrac{2\pi}{L}(\mathbf{e_1}+\mathbf{e_2})$&$D_{2h}$&$ B_1^-$&$ \mathbf{\rho_{1,2}}(\mathbf{e_1+e_2})\to \pi(\mathbf{e_1+e_2})\pi(\mathbf{0})$
\end{tabular}
\end{center}
\end{table*}

\vspace{-0.5cm}

\noindent
and all permutations in direction $\mathbf{P}$ and $\rho$-polarization. 
We display the symmetry group, the irreducible representation and the decay mode, which applies to three cases of $\mathbf{P}$ \cite{rummukainen,rho_etmc_pos,rho_etmc,rho_our}. 

Other simulations aimed at $\Gamma_\rho$ used at most one quark-antiquark interpolator and one $\pi\pi$ interpolator for each $\mathbf{P}$.  
We use 15 quark-antiquark interpolators ${\cal O}_{i=1-5}^{s=n,m,w}$ and one $\pi\pi$ interpolator for each $\mathbf{P}$, 
where each pion is projected to a definite momentum:
\begin{align}
\label{interpolators}
{\cal O}_{i=1,..,5}^{s}(t)&=\sum _{\mathbf{x}}\tfrac{1}{\sqrt{2}}~[\bar u_s(x) ~
{\cal F}_i ~\E^{\I \mathbf{Px}} ~u_s(x)\ -\bar d_s(x) ~
{\cal F}_i ~\E^{\I \mathbf{Px}} ~d_s(x)]\qquad 
(s=n,m,w) \FC\\
{\cal O}_6^n(t)&=\tfrac{1}{\sqrt{2}}
[\pi^+(\mathbf{p_1})\pi^-(\mathbf{p_2})-\pi^-(\mathbf{p_1})\pi^+(\mathbf{p_2})]
\ ,\qquad \pi^{\pm}(\mathbf{p_i})=\sum_{\mathbf{x}} \bar q_{n}(x)  \gamma_5 
\tau^{\pm} \E^{\I \mathbf{ p_i x}} q_{n}(x)\FD\nonumber
\end{align}
Quark-antiquark interpolators have five different color-spin-space structures ${\cal F}_i$. The quarks 
are smeared using the Laplacian Heaviside (LapH) smearing proposed in \cite{peardon}, i.e.,
\begin{equation}
\label{laph}
q_s\equiv \Theta(\sigma_s^2+\nabla^2)\;q=\sum_{k=1}^{N_v} 
\Theta(\sigma_s^2+\lambda^{(k)}) ~v^{(k)} v^{(k)\dagger}\,, \quad s=n~(narrow),~m~(middle),~w~(wide)\,,
\end{equation}
where different truncations $N_v=96,~64,~32$ correspond to three different widths $s=n,m,w$  \cite{rho_our}.

\begin{figure}[bt]
\begin{center}
\includegraphics*[width=0.5\textwidth,clip]{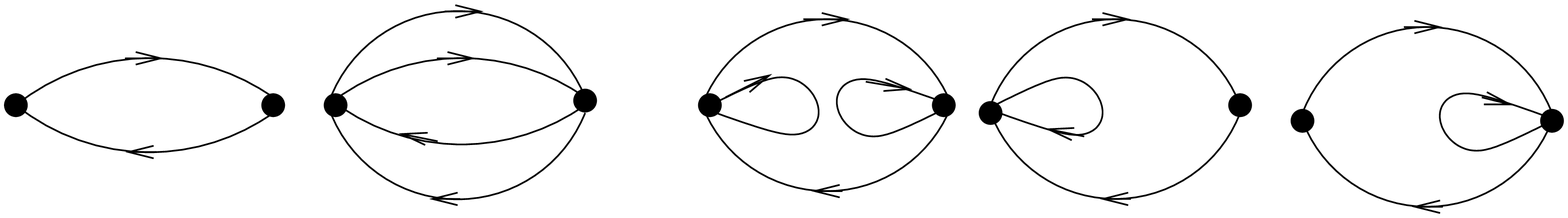}
\end{center}
\vspace{-0.5cm}
\caption{ Contractions for our $\rho$ and $a_1$ correlators with $\bar qq$ and $meson-meson$ interpolators ($I=1$).}\label{fig:contractions}
\end{figure}

The $16\times 16$  correlation matrix  $C_{ij}(t_f,t_i)=\langle 0|{\cal O}_i(t_f){\cal O}_j^\dagger (t_i)|0\rangle$ necessitates the inclusion of the contractions in Fig. \ref{fig:contractions}. 
The contractions were  computed using the full distillation method, which is based on   the LapH smeared quarks (\ref{laph}) \cite{peardon} and  leads to relatively precise results for all types of contractions in Fig. \ref{fig:contractions}. 
We apply this method for the first time to extract a meson width. We also propose how to apply it 
for  interpolators with different smearing widths in the same variational basis \cite{rho_our}. All correlators are expressed in terms of the so-called perambulators in Appendix A of \cite{rho_our}. The resulting correlators are averaged over all source time-slices $t_i$, over all directions  of $\mathbf{P}$ and $\rho$ polarization.

The lowest two energies of the system are determined using the 
Generalized Eigenvalue Method (GEVP) \cite{gevp} and the dependence on the choice of the interpolators in the variational basis is explored in \cite{rho_our}. 
The lowest energy level is robust to this choice. We find that the first excited energy level cannot be reliably obtained without the $\pi\pi$ interpolator in the basis, and that more than two interpolators are required at least in the case $P=\tfrac{2\pi}{L}(1,1,0)$. 
The extracted six energy levels for our preferred interpolator choice \cite{rho_our} are given in Table III of \cite{rho_our}. 

\begin{figure}[bt]
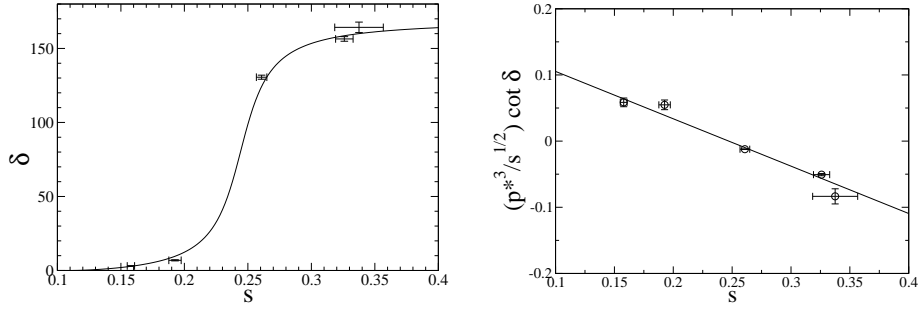

\begin{center}
\includegraphics[height=4.0cm,clip]{figs/phaseshift_and_fit.eps} $\quad$
\includegraphics[height=4.0cm,clip]{figs/BW_xy_fits.eps}
\end{center}

\vspace{-0.5cm}

\caption{ \small The phase shift $\delta$ (in degrees) for $\pi\pi$ scattering in P-wave and  $((ap^*)^3/\sqrt{sa^2}) \cot\delta$ as a function of $s$, together with a Breit-Wigner fit. }\label{fig:phaseshift_rho}
\end{figure}

Each of the six energy levels gives the value of the phase shift $\delta(s)$ at $s=E^2-\mathbf{P}^2$ (\footnote{We  use the discrete dispersion relation $cosh({\sqrt{s}a})=\cosh({Ea})-2 \sum_{k=1}^3 \sin^2(P_k a/2)$ instead of the continuum one $s=E^2-\mathbf{P}^2$ to analyze the $\rho$ \cite{rho_our,rho_cppacs}. We  analyze the $a_1$ using the continuum dispersion relation.  })
 via the L\"uscher formula for $\mathbf{P}=0$ or its generalization to $\mathbf{P}\not = 0$ \cite{rummukainen,rho_etmc_pos}. We  independently confirmed the needed relations and compiled them in \cite{rho_our}. One of these levels, $E_2(\mathbf{P}=0)$, is above the inelastic threshold $\sqrt{s}>4m_\pi$ and we omit it from further analysis.

The resulting phase shifts for five different values of $s$ are plotted 
in Fig.~\ref{fig:phaseshift_rho}. The phase shift has relatively small errors and  exhibits a resonating behavior, which allows us to extract $m_\rho$ and $\Gamma_\rho$ or rather the coupling $g_{\rho\pi\pi}$.  We use the Breit-Wigner relation (\ref{bw}) together with  $\Gamma(s)\equiv g_{\rho\pi\pi}^2 ~{p^*}^3/(6\pi s)$, which leads to 
\begin{equation}\label{linear_fit_rho}
\frac{{p^*}^3}{\sqrt{s}}\,\cot \delta(s)= \frac{6\pi}{g_{\rho\pi\pi}^2} ( m_\rho^2 - s )~.
\end{equation}
 This allows a linear fit in $s$ (Fig. \ref{fig:phaseshift_rho}) to extract  $m_\rho$ and $g_{\rho\pi\pi}$ given in Table \ref{tab:results_resonance}. The resulting $m_\rho a=0.4972(42)$ is slightly lower than the naive value $m_\rho^{naive} a=0.5107(40)$, which is extracted from the ground state  with $\mathbf{P}=0$. 
 We also extract the P-wave scattering length $a_{l=1}^{\pi\pi}=0.082(10)(3)~$fm$^3$ (defined as $a_{l}\equiv lim_{\delta\to 0} \delta(p^*)/{p^*}^{2l+1}$ \cite{a_scat_Pwave}) from the state with the lowest\footnote{The next state leads to $a_{l=1}^{\pi\pi}$ consistent with the value obtained from the lowest state.} $p^*a=0.1076(36)$ and $\delta=3.03(6)^\circ$. This qunatity is not directly measured, so we compare it to the typical value $a_{l=1}^{\pi\pi}\simeq 0.038(2)~(m_\pi^{phy})^{-3}$ obtained by combining experiment and ChPT or Roy equations \cite{a_scat_Pwave}.

A comparison of the resulting $m_\rho$ and $g_{\rho\pi\pi}$ to two recent lattice simulations \cite{rho_etmc,rho_pacscs} is compiled in \cite{rho_pacscs}. The $N_f=2$ simulation with twisted mass quarks \cite{rho_etmc} and the $N_f=2+1$ simulation with Wilson quarks were done at four/two values of $m_\pi$ and explicitly demonstrate the  mild dependence of $g_{\rho\pi\pi}$ on $m_\pi$. All three results on $g_{\rho\pi\pi}$ are relatively close to each other and close to the $g_{\rho\pi\pi}^{exp}=5.97$ extracted from $\Gamma_\rho^{exp}$. 
The resonance mass $m_\rho$ of \cite{rho_pacscs} is $\simeq 11\%$ higher than ours, while $m_\rho$ of \cite{rho_etmc} is $\simeq 21\%$ higher than ours, at comparable $m_\pi$. Note that all three simulations get the resonance $m_\rho $ within $3\%$ from the value of $m_\rho^{naive} $, which implies that the simulations differ already in $m_\rho^{naive} $. Possible causes for different $m_\rho^{naive} $ could be discretization effects or scale fixing of all three simulations, flavor breaking of twisted mass quarks  \cite{rho_etmc} or partial quenching of the strange quark  \cite{rho_etmc, rho_our}. 
Additional causes for the different $m_\rho $ values could be the
small interpolator basis in \cite{rho_etmc,rho_pacscs} or the small box $L\simeq 2$ fm 
in \cite{rho_our}. The exponentially suppressed terms, which are neglected in L\"uscher 
formulae, may not be completely negligible for our $L\simeq 2$ fm, 
which is a systematic uncertainty of our simulation. We are planing a simulation at larger $L$ to explore possible finite size effects. 
We believe, however, that our small $L$ does not influence our $m_\rho^{naive}$, as the first excited state $\pi(2\pi/L)\pi(-2\pi/L)$ at $\mathbf{P}=0$   hardly affects the $m_\rho^{naive}$ ground state.  

 Our $\delta(s)$ agrees reasonably well with the prediction of the lowest\footnote{One cannot make a fair comparison between our lattice result and the NLO prediction, since it depends on a number of LECs, and some of them have been fixed using $m_\rho$ from another lattice study, which gets a significantly higher $m_\rho$.}  order of Unitarized Chiral Perturbation Theory \cite{pelaez_comparison}, which has been recalculated for our $m_\pi=266$ MeV. 
\begin{table*}[t]
\begin{center}
\begin{tabular}{c | c c  c | c c  c c}
     &   $m_\rho~$[MeV]  & $g_{\rho\pi\pi}$ & $a_{l=1}^{\pi\pi}$ & $m_{a1}~$[GeV] & $g_{a_1\rho\pi}~$[GeV] & $a_{l=0}^{\rho\pi}~$[fm] &\\
\hline
 latt  &  $792(7)(8)$ & $5.13(20)$ & $0.082(10)(3)$ & $1.44(4)$   & $1.1(3)$ & $0.23(12)$ & {\small using} $m_\rho$ \\  
&   & & & $1.43(5)$  & $1.7(4)$ & $0.56(23)$ & {\small using} $m_\rho^{naive}$ \\ 
\hline
exp & $775.5$  & $5.97$ & $0.108(5)~$* & $1.23(4)$  & $<1.35(30)$  & {\small not meas.} & \\  
 \end{tabular}
\caption{Our lattice results for the resonance properties \cite{rho_our}, compared to the experimental values. The results related to $a_1$ depend on the choice of the input $\rho$ mass: $m_\rho$ or $m_\rho^{naive}$. The experimental value of $a_{l=1}^{\pi\pi}~$ is obtained combining experiment with ChPT or Roy equations.\label{tab:results_resonance}  
 }
\end{center}
\end{table*}

\section{The $\rho\pi$ phase shift and $a_1$ resonance}

We study the S-wave scattering of $\rho\pi$, where the resonance $a_1(1260)$ appears, 
  for the total momentum $\mathbf{P}=0$.  The scattering is elastic at least until   $a_1(1260)$ on our lattice since $\bar K^*K$ cannot be created on our $N_f=2$ ensemble. 
The ground scattering state  is $\rho(\mathbf{0})\pi(\mathbf{0})$  in the non-interacting limit.
The scattering particle $\rho(\mathbf{0})$  is almost stable on our lattice, since its lowest decay channel $\pi(2\pi/L)\pi(-2\pi/L)$ is significantly higher in energy. 

We use 9 quark-antiquark interpolators ${\cal O}_{i=1-3}^{s=n,m,w}$ and one 
$\rho(\mathbf{0})\pi(\mathbf{0})$ interpolator, all with $J^{PC}=1^{++}$, 
$|I,I_3\rangle =|1,0\rangle$ and $\mathbf{P}=0$:
\begin{align}
\label{interpolators_a1}
{\cal O}_{1}^s(t)&=\sum _{\mathbf{x},i}\tfrac{1}{\sqrt{2}}~\bar u_s(x) ~
A_i\gamma_i~\gamma_5~ \E^{\I \mathbf{Px}} ~u_s(x)\ -\{u_s\leftrightarrow d_s\}\qquad 
(s=n,m,w) \FC\\
{\cal O}_{2}^s(t)&=\sum_{\mathbf{x},i,j}\tfrac{1}{\sqrt{2}}~\bar u_s(x)
\overleftarrow{\nabla}_j~A_i\gamma_i~\gamma_5~ \E^{\I \mathbf{Px}}~
\overrightarrow{\nabla}_j u_s(x)\ - \{u_s\leftrightarrow d_s\}\qquad(s=n,m,w)\FC\nonumber\\
{\cal O}_{3}^s(t)&=\sum_{\mathbf{x},i,j,k} \tfrac{1}{\sqrt{2}}~\epsilon_{ijl} 
 ~\bar u_s(x)~A_i\gamma_j ~\tfrac{1}{2}[\E^{\I \mathbf{Px}} 
 \overrightarrow{\nabla}_l  -\overleftarrow{\nabla}_l  \E^{\I \mathbf{Px}}] 
  u_s(x) -\{u_s\leftrightarrow d_s\}\quad(s=n,m,w)\FC\nonumber\\
{\cal O}_4^n(t)&=\tfrac{1}{\sqrt{2}}
[\pi^+(\mathbf{0})\rho^-(\mathbf{0})-\pi^-(\mathbf{0})\rho^+(\mathbf{0})]
\; ,\quad \pi^{\pm}(\mathbf{0})=\sum_{\mathbf{x}} \bar q_{n}  \gamma_5 
\tau^{\pm}q_{n}~,\quad \rho^{\pm}(\mathbf{0})=\sum_{\mathbf{x}} \bar q_{n}  A_i\gamma_i 
\tau^{\pm}q_{n} \;,\nonumber
\end{align}
where $\nabla$ is the covariant derivative. 
The contractions in Fig.~\ref{fig:contractions} are calculated using the full distillation method 
and averaged over all source time slices $t_i$ and all $a_1$ polarizations $\mathbf{A}$. 

\begin{figure}[bt]
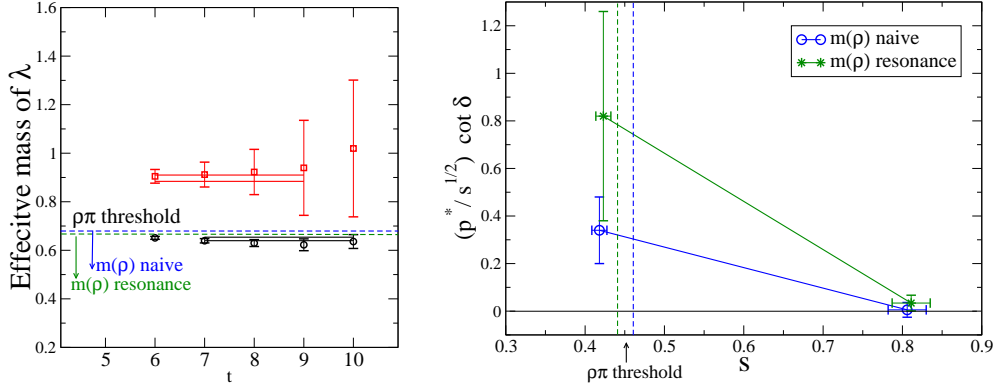

\begin{center}
\includegraphics[height=5.1cm,clip]{figs/eff_a1_lat.eps}
$\quad$ 
\includegraphics[height=5.1cm,clip]{figs/BW_xy_a1.eps}
\end{center}

\vspace{-0.5cm}

\caption{ \small The effective mass for lowest two eigenvalues in $a_1$ channel (left). The  combination $p^* \cot\delta/\sqrt{s}$ as a function of $s=E^2-\mathbf{P}^2$, where $\delta$ is $\rho\pi$ phase shift in S-wave (right).}\label{fig:phaseshift_a1}
\end{figure}

The effective mass for the lowest two eigenvalues  are shown in Fig. \ref{fig:phaseshift_a1} and the  resulting  $E$ and $p^*$ are given in Table \ref{tab:res_a1}.  The CMF momentum $p^*$ is extracted using $E=\sqrt{p^{*2}+m_\pi^2}+\sqrt{p^{*2}+m_\rho^2}$:  it is imaginary for the ground state below $m_\pi+m_\rho$ threshold, and real for the first excited state. 
We take two choices for the input $\rho$ mass: our main results are based on the resonance mass $m_\rho$  
(green lines in Fig. \ref{fig:phaseshift_a1}), while $m_\rho^{naive}$ is taken for comparison (blue lines in Fig. \ref{fig:phaseshift_a1}). 

The S-wave phase shift $\delta$ for $\mathbf{P}=0$  is extracted using the well known L\"uscher relation \cite{luscher}
\begin{equation}
\label{luscher_a1}
p^* \cot\delta=\frac{2}{\sqrt{\pi} ~L}~Z_{00}\bigl(1,(\small{\tfrac{p^* L}{2\pi}})^2\bigr)  \stackrel{p^*\to 0}{\longrightarrow} \frac{1}{a^{\rho\pi}_{l=0}}\ ,
\end{equation}
which applies above and below threshold. The results are compiled in Table \ref{tab:res_a1} for both choices of $\rho$ mass. The first excited level gives $\delta \approx 90^\circ$, so it is sitting close to the top of the $a_1$ resonance and  $m_{a1}\approx E_2$ holds.   The ground state with imaginary $p^*$ gives imaginary $\delta$, but the product $p^* \cot\delta$ is real since $Z_{00}(1,(\frac{p^* L}{2\pi})^2)$ is real.  

We parametrize $\Gamma_{a1}(s)\equiv g_{a_1\rho\pi}^2 ~p^* /s$ and apply the Breit-Wigner relation (\ref{bw}) to get 
\begin{equation}
\label{linear_fit_a1}
\frac{{p^*}}{\sqrt{s}}\,\cot \delta(s)= \frac{1}{g_{a_1\rho\pi}^2} ( m_{a1}^2 - s )~,
\end{equation}
which applies in the vicinity of the resonance above or below threshold. 
Given the values of $p^* \cot\delta$ at two different values of $s$,
we apply a linear fit (\ref{linear_fit_a1}) in $s$ (shown in Fig. \ref{fig:phaseshift_a1})  to extract $m_{a1}$ and $g_{a_1\rho\pi}$. The results are compiled  in Table \ref{tab:results_resonance}. 
Our $m_{a1}$ at $m_\pi=266$ MeV is about $14\%$ higher than the experimental resonance $a_1(1260)$. The first lattice result for $g_{a_1\rho\pi}$ is  valuable, since this coupling is not known experimentally. None of the $a_1$ branching ratios have been measured, so we provide only the upper limit for $g^{exp}_{a_1\rho\pi}$  resulting from  the total width $\Gamma^{exp}_{a1}=250-600$ MeV. Our lattice result $g_{a_1\rho\pi}=1.1(3)$ GeV is in agreement with the value $g_{a_1\rho\pi}^{phen}\approx 0.9$ GeV obtained using Unitarized Effective Field Theory approach \cite{axial_oset} and converted to our convention. We extract also $a_{l=0}^{\rho\pi}$ from the ground state, which is sufficiently close to the threshold. The scattering experiment cannot be carried out since $\rho$ is a quickly decaying  particle, so we  compare our $a_{l=0}^{\rho\pi}({\small m_\pi\!=\!266~MeV})=0.23(12)$ fm  to $a_{l=0}^{\rho\pi}(m_\pi^{phy})\approx 0.37$ fm  obtained from Unitarized Effective Field Theory  \cite{oset_roca_private}.

\begin{table*}[t]
\begin{tabular}{ccccccc}
 level &  fit  &   $Ea=\sqrt{s}a$ 
& $p^*a$ &  $\delta$ & $p^*cos(\delta)/\sqrt{s}$ &\\
\hline
1&   7-10& $0.6468(73)$& $\I~0.065(13)$  & $\I~ 7.1(54)^\circ$  & $0.82(44)$ & (using $m_\rho$)  \\
 & & & $\I~0.086(9)$ & $\I~ 23(14)^\circ$ & $0.34(14)$ & (using $m_\rho^{naive}$) \\
\hline
2&   6-9 &$0.897(13)$& $0.280(10)$    & $83.7(59)^\circ$  & $0.034(33)$ & (using $m_\rho$)   \\
 & & & $0.272(10)$ & $88.9(59)^\circ$ & $0.005(31)$ & (using $m_\rho^{naive}$)\\
\hline
\end{tabular}
\caption{\label{tab:res_a1}
The results for the $a_1\leftrightarrow \pi\rho$ coupled channel with interpolators ${\cal O}_{1,2,4}^n$ and GEVP reference time $t_0=5$. 
The ground state is below $\rho\pi$ threshold, so $p^*$ and $\delta$ are imaginary, while the $p^*\cot\delta$ is real.   }
\end{table*}

\section{Conclusions}

The lattice extraction of the phase shifts for elastic scattering has recently 
become  possible also for the attractive resonant channels. We simulated the 
scattering in the $\rho$ and $a_1$ channels and extracted the mass and the width of these two resonances 
as well as the scattering lengths in the corresponding meson-meson channels.  

\acknowledgments
\noindent
 
We would like to thank  A. Hasenfratz for providing the gauge configurations used for this work and A. Rusetsky for valuable discussions related to the $a_1$ channel.  We would also like to thank G. Colangelo, G. Engel, X. Feng, N. Ishizuka, E. Oset, L. Roca,  G. Schierholz and R. Woloshyn for helpful discussions. This work is
supported by ARRS and by the Natural Sciences and Engineering
Research Council of Canada (NSERC).

\end{document}